**Article information**

**Article title**

*Impact Range Assessment (IRA): An Interpretable Sensitivity Measure for Machine Learning Regression Modelling*

**Authors**

*Jihao You[1]*, Dan Tulpan[1], Jiaojiao Diao[2], Jennifer L. Ellis[1]*

**Affiliations**

*[1] Department of Animal Biosciences, University of Guelph, Guelph, ON, Canada*
*[2] Department of Integrative Biology, University of Guelph, Guelph, ON, Canada*

**Corresponding author's email address and Twitter handle**

**jyou03@uoguelph.ca*

**Abstract**

While machine learning regression models capture the relationship between predictors and the response variable, they often lack intuitive methods to help modelers understand how sensitive model predictions are to changes in individual predictors. To address this limitation, we introduce a model-agnostic interpretability method called Impact Range Assessment (IRA), which quantifies the sensitivity of model predictions to changes in each predictor by measuring the range of predicted responses as the predictor varies across its observed range. Validation using synthetic linear and nonlinear datasets demonstrates that relevant predictors produced higher IRA values than irrelevant ones. Moreover, repeated evaluations produced results aligned with those from the single-execution analysis, confirming the method's robustness. For demonstration, two benchmark datasets were used to evaluate IRA across five machine learning regression models, yielding model-specific sensitivity rankings and supporting IRA's applicability as a complementary sensitivity analysis tool alongside feature importance methods such as permutation feature importance and SHapley Additive exPlanations (SHAP).

- IRA quantifies (for regression models) the sensitivity of regression model predictions to changes in each predictor across its observed range.
- IRA is model-agnostic and applicable to a wide range of linear and nonlinear regression models.
- IRA supports fast single-execution analysis or repeated analysis to assess result stability.

**Graphical abstract**

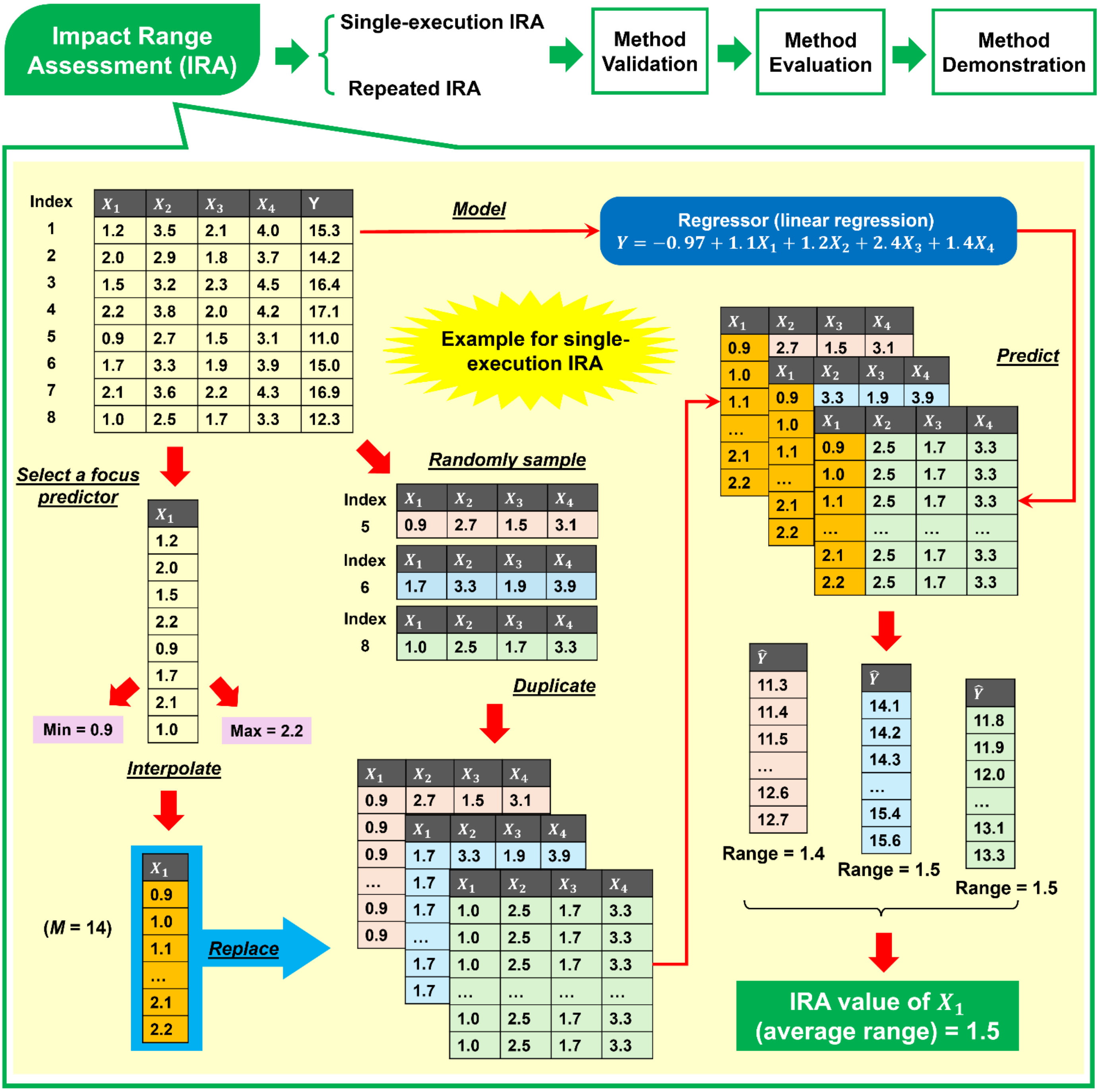

## Specifications table

| | |
|---|---|
| **Subject area** | Computer Science |
| **More specific subject area** | *Model interpretability approaches for regression models* |
| **Name of your method** | *Impact Range Assessment (IRA)* |
| **Name and reference of original method** | *N.A.* |
| **Resource availability** | *The Python script for the IRA functions is available at the GitHub link: https://github.com/jyou2025/impact-range-assessment. The two synthetic datasets were created using the scikit-learn library, and the California Housing dataset was also obtained from the scikit-learn library. The Concrete Compressive Strength dataset was obtained from UC Irvine Machine Learning Repository. The dataset for the PDI prediction case study is confidential.* |

**Background**

Regression methods model the relationship between multiple predictors and a continuous numeric response variable and are widely used for prediction and estimation [1]. When constructing regression models, it is essential to understand how sensitive model predictions are to changes in individual predictors. The general problem is that for linear regression models, coefficients cannot be compared across predictors unless the predictors share the same scale. Once predictor values are standardized, however, the coefficients no longer directly represent changes in the predicted response per unit change in the original predictors. For nonlinear regression models, especially machine learning models, the comparison becomes even more difficult as most models typically rely on complex structures that lack easily interpretable parameters.

In machine learning, feature importance methods such as Permutation Feature Importance [2] and SHAP [3] are widely used to quantify the predictive strength of individual variables. However, these methods primarily focus on isolating each predictor's contribution to model predictions. To better capture the dynamic behavior of a model, sensitivity analysis provides a more appropriate framework for determining how changes in model inputs contribute to variations in model output [4]. Derivative-based local approaches to sensitivity analysis assess changes in the response variable by perturbing a single input around a baseline value [5]. However, this local approach only captures effects within a specified region of the input space (for example, +/-25 % around the mean), without providing explicit means to explore the full input space [6]. Compared to local sensitivity analysis, global sensitivity analysis provides an alternative and more robust sensitivity measure, and considers nonlinearity by sampling from a wider parameter space [7]. Global sensitivity analysis includes various approaches, such as variance-based methods like the Sobol' method [8], and difference-based methods like the Morris method [9]. Since global sensitivity analysis methods evaluate the impact of predictors through mathematical formulations, they are often computationally intensive and difficult to interpret and are primarily applied in mathematical simulation models rather than data-driven machine learning regression models. Visual sensitivity analysis approaches such as Partial Dependence Plots (PDP) [10] and Individual Conditional Expectation (ICE) [11] plots offer intuitive representations of how individual predictors affect model output. The PDP shows the average effect of each predictor on the response variable, while ICE plots reveal how individual observations respond differently. Scatter plots similarly provide qualitative visual indications of relationships between predictors and the observed response variable [12]. However, such visual inspection approaches lack a straightforward way to quantify these relationships or compare them across predictors. To bridge this gap, there is a need for an intuitive and efficient approach to quantify and compare the sensitivity of model predictions to changes in individual predictors across their observed range.

To address these limitations, we propose a new sensitivity measure named the Impact Range Assessment (**IRA**), which evaluates the potential change in a regression model's response variable when each numeric predictor is varied across its observed range. IRA offers great flexibility: it can be deployed as a single-execution procedure for a quick estimation of predictor impact or executed repeatedly to assess the stability of the results. Unlike conventional sensitivity methods, IRA provides a straightforward quantitative basis for directly comparing and ranking predictor effects, offering a way to interpret and rank the impact of predictors on the response variable.

## Method details

The IRA quantifies the sensitivity of model predictions to changes in individual predictors by measuring the range of predicted responses as each predictor is varied across its observed range. The resulting score represents the potential change in the model output, simultaneously accounting for the model's internal sensitivity to that predictor and the distribution of the input data. Implementing the IRA approach requires only a trained regression model and its corresponding dataset containing continuous numeric predictors.

The pseudocode for the IRA method is shown in Figure 1 and a linear regression model built on a simple dataset with eight observations and five variables (four predictors and one response variable) from the Graphical Abstract is used to illustrate the algorithm. The regression coefficients, predicted values, and IRA ranges shown in the Graphical Abstract are rounded.

The IRA method consists of five steps:

1. *Select a focus predictor.* Isolate a single predictor of interest ($X_f$). Each predictor in the dataset is evaluated individually by treating it as the focus predictor ($X_1$ in the Graphical Abstract).
2. *Interpolate values across the predictor's range.* Generate $M$ evenly spaced $X_f$ values spanning from the predictor's minimum to maximum values. For example, a variable $X_1$ ranging from 0.9 to 2.2 with a step of 0.1 produces 14 interpolated points ($M$ =14).
3. *Generate background observations.* Randomly draw $K$ observations from the input dataset, using sampling with replacement. In the example, three observations ($K$ = 3) corresponding to indices 5, 6, and 8 are selected.

4. *Create modified input sets.* Replicate each of the $K$ background observations *M* times. In each replicated set, replace the original values of the focus predictor with the $M$ interpolated points while keeping all the other predictors constant. In our example, each sampled observation was replicated 14 times, resulting in three modified input sets with updated $X_1$ values.
5. *Calculate the IRA value.* Use the trained regression model to predict the response variable for all modified input sets. For each background observation, determine the minimum and maximum predicted values across its $M$ variations. The range for each background observation, defined as the difference between its minimum and maximum values, is computed. The final IRA value for the focus predictor is the average of these $K$ ranges. In our example, the predicted ranges of the three modified input sets (*K* = 3) were 1.4, 1.5, and 1.5, yielding a final IRA value of 1.5 for $X_1$.

The IRA equation is shown below:

$$IRA_i = \frac{1}{K}\sum_{k=1}^{K}[max(\hat{y}_i^k) - \min(\hat{y}_i^k)]$$

where $i$ is the $i^{\text{th}}$ predictor, *K* represents the number of background observations, and $\max(\hat{y}_i^k)$ and $\min(\hat{y}_i^k)$ are the maximum and minimum predicted outputs for the $k^{\text{th}}$ background observation. Because IRA quantifies the change in predicted output values, its unit is identical to that of the output variable.

The IRA values are always non-negative, where a higher value indicates a greater sensitivity of model predictions to changes in the corresponding predictors. The IRA values can be affected by two main hyperparameters: (1) the number of interpolated points ($M$) for each focus predictor, which determines the resolution of the predictor's range, and (2) the number of background observations ($K$) randomly drawn with replacement, which defines the size of the background dataset. The computational complexity of IRA scales as a function of four variables: the number of predictors ($p$), the number of interpolated points ($M$), the number of background observations ($K$), and the inference time of the underlying model ($T$). The overall computational cost is approximately proportional to $p \times M \times K \times T$. In practice, increasing *M* has minimal impact on runtime since predictions can be efficiently batched. $T$ may dominate the total runtime for computationally intensive models.

To quantify the statistical uncertainty associated with random resampling of background observations, a repeated IRA procedure is also proposed to generate confidence intervals (CIs) for each predictor. In this procedure, the single-execution IRA is repeated several times, with each repeat generating an IRA value based on a unique resampling of background observations. This iterative process generates an empirical distribution of IRA values per predictor from which 95% CIs are derived using the 2.5th and the 97.5th percentiles [13].

## Method implementation

All data analyses were conducted in Python 3.12 using scikit-learn 1.8.0 [14], NumPy 2.2.5 [15], pandas 2.3.3 [16], and Matplotlib 3.10.9 [17].

### *1. Method validation*

To evaluate the proposed IRA method, two synthetic datasets were generated to represent different relationships between predictors and the response variable, allowing for a controlled assessment of the framework with varying levels of model complexity.

- Linear Dataset: Generated using the *make_regression()* function with additive bias and noise. It contains 5,000 samples and eight predictors ($X_1$, $X_2$, … $X_8$), with values drawn from a normal distribution $\mathcal{N}(0, 1^2)$. Five predictors ($X_2$, $X_3$, $X_4$, $X_6$ and $X_8$) were designated as relevant predictors and contributed to the target variable $Y$, while the remaining three ($X_1$, $X_5$ and $X_7$) served as irrelevant predictors. Detailed descriptive statistics are provided in Table S1.
- Nonlinear Dataset: Generated using the *make_friedman1()* function from the *scikit-learn* library, a widely used nonlinear benchmark featuring nonlinear interactions [18]. The dataset has 5,000 samples and eight predictors ($X_1$, $X_2$, … $X_8$), with all values drawn from a uniform distribution $\mathcal{U}(0,1)$. Based on the Friedman benchmark definition [18], the first five predictors ($X_1$, $X_2$, $X_3$, $X_4$ and $X_5$) contributed to the nonlinear response, while the remaining three ($X_6$, $X_7$ and $X_8$) were added noise predictors. The descriptive statistics of the nonlinear dataset are provided in Table S2.

Both synthetic datasets were used to evaluate hyperparameter sensitivity for the single-execution IRA, across different interpolation resolutions ($M \in \{50, 100, 200, 500\}$) and background sample sizes ($K \in \{50, 100, 200, 500\}$). Additionally, to assess the stability of the bootstrap estimates, repeated IRA analyses were performed with iterations ranging from 10 to 90 (in increments of 20). Ordinary Least Squares using the *LinearRegression()* function and a Random Forest (RF) regressor using the *RandomForestRegressor()* function from the scikit-learn library were applied to the linear and nonlinear datasets, respectively.

### *2. Method evaluation*

To evaluate the generalizability of the IRA method across different architectural paradigms, two public benchmarks were used.

- California Housing Dataset: Introduced by Pace and Barry [19] and available through the *scikit-learn* library [14], this dataset contains 20,640 observations and eight numeric predictors derived from the 1990 U.S. Census. The response variable is the *Median House*

*Value* for California districts (expressed in units of 100,000 USD), predicted using eight localized demographic and geographic features (Table S3).

- Concrete Compressive Strength Dataset: Obtained from UC Irvine Machine Learning Repository [20], the dataset contains 1,030 concrete formulations. The prediction target is the *Compressive Strength (MPa)* of concrete, modeled via eight numeric predictors characterizing material mixture components and curing age (Table S4).

Five machine learning regressors were evaluated: Gradient Boosting (GB), Multilayer Perceptron (MLP), RF, Ridge Regression (RR), and Support Vector Regression (SVR). All models were implemented using functions from the *scikit-learn* library [14]. All models were implemented with default hyperparameters, except for the MLP model, which was trained with a maximum of 500 iterations to ensure convergence. As the goal was to demonstrate IRA's applicability across different model types, all models were fitted on the full dataset without a train-test split. A fixed random seed was applied across all stochastic algorithms to ensure reproducibility. Prior to fitting, inputs for scale-sensitive models (MLP, RR and SVR) were normalized using the *StandardScaler()* function from the *scikit-learn* library [14]. GB and RF were trained on the unscaled data.

To contextualize the IRA results, the repeated IRA results were compared against established interpretability methods: Permutation Feature Importance (using Mean Absolute Error as the loss measure to assess performance drops [2]), mean absolute SHAP (SHapley Additive exPlanations) values to quantify the magnitude of feature contributions, and standard Partial Dependence Plots (PDP) and Individual Conditional Expectation (ICE) curves.

### *3. Method Demonstration*

The IRA approach was applied to a real-world agricultural engineering case study using data from You et al. [21]. The objective of this study was to evaluate manufacturing features that influence animal pelleted feed quality. A multiple linear regression model was constructed to predict the Pellet Durability Index (PDI, a widely used measure for quantifying the quality of pelleted feed), using nine manufacturing predictors (Table S5) across 2,149 process observations. Following the methodology of You et al. [21], the response variable was transformed via a Box-Cox transformation (tPDI) to correct for skewness in the empirical PDI distribution, and predictions from the resulting model were eventually back-transformed to the original PDI scale. This transformation ensured that while a linear model structure was maintained, underlying system nonlinearities were effectively accommodated.

## Results

### *1. Method validation*

The single-execution IRA clearly distinguished relevant from irrelevant predictors in the LR model (Table 1). Predictors that contributed directly to the response variable ($X_2$, $X_3$, $X_4$, $X_6$ and $X_8$) consistently exhibited substantially higher IRA values than the irrelevant noise variables ($X_1$, $X_5$ and $X_7$). For example, $X_2$ had an IRA value of 632.92, indicating that varying this feature across its observed min to max range (-3.837 to 3.728) shifted the predicted response by 632.92 units. In contrast, the irrelevant predictor $X_1$ shifted the response by a maximum of only 5.69 units. Moreover, because the computed IRA values remained identical across all combinations of interpolation points ($M$) and background sample sizes ($K$), the repeated IRA was not required for the LR model.

For the RF model, the single-execution IRA showed that the relevant predictors ($X_1$, $X_2$, $X_3$, $X_4$ and $X_5$) consistently exhibited higher IRA values compared to the irrelevant ones ($X_6$, $X_7$ and $X_8$) (Table 2), aligning with their nonlinear contributions to the outcome. For example, the relevant predictor $X_4$ had an IRA value range of 11.53–12.36, indicating a ~12-unit shift in the response variable when $X_4$ varied between its minimum (0) and maximum (1) bounds. In contrast, the irrelevant predictors ($X_6$, $X_7$ and $X_8$) affected the response variable by up to approximately 2.75 units.

When applying the repeated IRA procedure to the Random Forest regressor across various bootstrap iterations, the mean IRA values remained stable (Table 3) and closely matched those obtained with the single-execution IRA (Table 2). The 95% confidence intervals (CIs) around the mean IRA values were relatively narrow for all predictors. Additionally, both the mean IRA values and confidence interval ranges showed only minor fluctuations as the number of repetitions increased from 10 to 90, with most predictors reaching stable parameter estimates at approximately 30–50 iterations.

***2. Method evaluation***

The repeated IRA framework was evaluated on the California Housing dataset across five regression models: GB, MLP, RF, RR, and SVR (Table 4). *Median Income* exhibited the largest IRA value in GB and RF, with predicted *Median House Value* spanning an average of 3.08 (GB) and 3.15 (RF) units change. This corresponds to an approximate 300,000 USD shift in localized home value. *Longitude* ranked first in the SVR model, with an average shift of 406,000 USD. *Average Rooms* had the highest IRA value in the MLP model, yielding an average shift of 3,448,000 USD in *Median House Value*, whereas *Average Bedrooms* ranked first in the RR model, followed by *Average Rooms*. In contrast, *House Age* consistently had the lowest IRA values in the MLP and SVR models, accounting for an average predicted change of 58,000 to 69,000 USD in *Median House Value. Population* and *Average Bedrooms* had the lowest IRA values in the GB and RF models, respectively.

As summarized in Table 5, *Cement* and *Age* consistently ranked among the predictors with the highest IRA values across all five models for the Concrete Compressive Strength dataset. *Cement* produced the largest overall IRA value, where variation across its observed range resulted in an average compressive strength change of 37.99 MPa (GB), 53.61 MPa (MLP), 32.31 MPa (RF), and 51.72 MPa (RR). *Age* ranked second in the GB, RF, and RR models (with IRA values of 36.47

MPa, 30.79 MPa, 41.49 MPa, respectively), and third in the MLP model. In contrast, for the SVR model, *Age* ranked first with an IRA value of 21.27 MPa, followed by *Cement* at 18.48 MPa. *Fly Ash* consistently had the lowest IRA values in the GB, RF and SVR models, while *Coarse Aggregate* showed the lowest IRA value in the RR model.

***3. Method Demonstration***

The repeated IRA was subsequently applied to a real-world multiple linear regression model predicting PDI from a feed mill dataset. The results in Table 6 show that *Fat Content* and *ADF Content* consistently exhibited higher IRA values than *Ambient Humidity* and *Cumulative Production*, indicating greater sensitivity of model predictions to changes in these predictors. Additionally, the repeated analysis produced relatively narrow 95% confidence intervals (CIs) across all features, indicating low variability and strong stability in the resulting IRA estimates.

## Discussion

***1. Parameter selection for the IRA framework***

Based on the results of the model validation, it is clear that the single-execution IRA is insensitive to internal parameter configurations for linear regression models. In contrast, although the IRA values for nonlinear regression models exhibited some fluctuations in the single-execution IRA with different parameter settings (i.e., the number of interpolated points and the number of background observations), the overall IRA values remained stable. Regarding the interpolation resolution ($M$), using only 50 points slightly underestimated IRA values for predictors involved in nonlinear relationships. Increasing the resolution to $M \geq 100$ resulted in more stable IRA estimates, while further increases caused only relatively small changes. Similarly, increasing the number of background observations ($K$) from 50 to 100 noticeably changed IRA estimates for several predictors, whereas additional increases had negligible effects. Overall, a configuration of $M$=100 interpolated points and $K$=200 background observations provided stable and reliable IRA estimates for this nonlinear model (Random Forest regressor). Although larger numbers of interpolated points and background observations may further improve IRA estimation for complex nonlinear models, it would also substantially increase computational cost.

Compared to the single-execution IRA that provides only point estimates, the repeated IRA procedure yields both mean IRA values and 95% confidence intervals (CIs), capturing variability from background predictor sampling and providing a more robust measure of the estimated impact for each predictor. The repeated IRA estimates showed only minor changes in IRA estimates as the number of iterations increased, showing negligible variance beyond 50 repeats. Consequently, executing 50 repeats provided an effective balance between statistical stability and computational efficiency for this nonlinear workflow. However, the appropriate number of repeats may vary across datasets and models and should be assessed for each case. As shown in Table S6, most models completed a single IRA execution in under one second. SVR was a notable exception, requiring 10.612 and 167.182 seconds on the Concrete Compressive Strength and California Housing datasets, respectively, which reflects SVR's higher prediction cost rather than structural overhead from the IRA algorithm itself.

***2. Influence of model architecture on IRA results***

Across both benchmark datasets, individual predictors exhibited different IRA values across different model architectures. For example, in the California Housing example, *Median Income* showed an average IRA value of 3.08 when using GB, compared to 2.71 (MLP), 3.15 (RF), 6.33 (RR), and 3.77 (SVR). These differences directly reflect how each model architecture responds to predictor variations. RR, a linear regression model, estimates predictor effects directly through model coefficients, with the range of predicted responses determined by the magnitude of each coefficient and the corresponding predictor's observed range. Tree-based models such as GB and RF partition the predictor space through ensemble-based decision structures, while SVR relies on kernel-based margins and support vectors to model nonlinear relationships. MLP models capture predictor effects through multiple layers of weighted connections and nonlinear activation functions.

Together, these patterns demonstrate that IRA values are jointly impacted by model architecture, feature ranges, and input distributions. Additionally, due to their intrinsic sensitivity profiles and computational mechanisms, some models might exhibit higher IRA values in the presence of extreme feature values or long-tailed distributions. For example, extreme values were present in *Average Rooms* in the California Housing dataset and in *Blast Furnace Slag* in the Concrete Compressive Strength dataset. Compared to GB, RF, and SVR, both the MLP and RR models exhibited disproportionately higher IRA values for these skewed predictors. For RR, this might be attributed to the combination of relatively large regression coefficients and wide predictor ranges, whereas for MLP, nonlinear transformations across multiple layers might produce substantial changes in model predictions when predictors are varied toward the extremes of their observed ranges. These findings also highlight the potential effects of extreme predictor values and potentially unrealistic combinations of input features on IRA estimates. While IRA measures model sensitivity over the observed predictor ranges, a large IRA value does not necessarily imply a realistic change in the response variable.

***3. Comparison with other feature explanation approaches***

Traditional machine learning interpretability tools, such as permutation feature importance and SHAP, explore the relative contributions of individual predictors to model predictions. Permutation feature importance quantifies the reduction in prediction performance caused by permuting the values of a predictor, while SHAP quantifies the contribution of a predictor to the model prediction relative to a baseline prediction. Sensitivity analysis, however, as explored with IRA, addresses a fundamentally different question: how the magnitude of the model’s output changes when inputs are systematically perturbed across an empirical range. Ranking features based on IRA identifies the model inputs to which predictions are most sensitive across the observed data range.

Figure S1 and Figure S2 show that permutation feature importance and SHAP exhibited strong agreement in feature ranking across all models, whereas IRA produced distinct rankings. This difference suggested that IRA captures a distinct but complementary aspect of predictor influence related to output sensitivity rather than their contribution to prediction accuracy. Classical global sensitivity analysis methods, such as Sobol' variance decomposition or Morris screening, are primarily designed and optimized for simulation-based mathematical models and often require extensive sampling procedures and variance decomposition frameworks. In contrast, IRA was designed as an interpretable, model-agnostic and computationally straightforward sensitivity-oriented approach tailored specifically for fitted regression models, particularly machine learning models, trained on observational datasets.

Existing visual sensitivity analytic approaches, such as PDP and ICE plots (Figure S3 and Figure S4), show individual conditional expectations (blue lines) and average partial dependencies (red lines). While PDP and ICE plots visually illustrate the directional relationship between predictors and outputs, they lack a straightforward quantitative basis for comparing predictor effects. The proposed IRA fills this gap for machine learning regression models by quantifying the variation of output values in response to variations of a given predictor's values. Unlike PDP and ICE, IRA produces a single numeric value per predictor, making it straightforward to compare and rank predictors directly. Therefore, the IRA serves as a complementary tool alongside feature importance, providing a clear and real-world relevant understanding of each predictor's individual impact on the model's predictions.

***4. Implications***

As a model-agnostic approach, the IRA provides significant practical advantages for interpreting complex regression models. First, IRA offers an easy way to interpret results with clear, real-world relevance, as it provides an intuitive way to assess predictor impacts that are accessible to both technical and non-technical users. This clarity is particularly important for interpreting complex models, such as machine learning models, as well as linear regression models with predictors on different scales or transformed variables. Since IRA values are expressed in the same unit as the model's response variable, the resulting measures translate directly into tangible, domain-specific insights. Second, the IRA approach establishes a rigorous quantitative basis for ranking variables based on the estimated IRA values. For a regression model built on a given dataset, model predictions are more sensitive to changes in predictors with higher IRA values, indicating larger variations in predicted responses when these predictors are varied across their observed ranges. This provides a quantitative basis for predictor ranking and prioritization. This ranking capability is particularly valuable in applications where understanding the relationships between predictors and the response variables is critical, which can support model interpretability and actionable insights across a wide range of disciplines. For example, the predictors in the PDI prediction case study could be straightforwardly ranked based on their IRA values. Higher IRA values for predictors such as *Fat Content* and *ADF Content* indicate greater sensitivity of model predictions to changes in these variables, reflecting their potential to produce larger variations in predicted PDI values and suggesting that they may require greater attention during pellet feed manufacturing. Therefore, in this case study, the IRA approach may help feed mill operators better understand the critical points in the pellet production process, and more broadly the IRA approach also helps data scientists identify predictors to which model predictions are most sensitive.

***5. Limitations***

Although the IRA framework offers an intuitive, model-agnostic way to assess how changes in individual predictor variables affect the magnitude of the response variable, several limitations should be considered when applying the method. First, IRA is primarily designed for continuous numeric predictors and discrete numeric predictors for which interpolation is mathematically meaningful. It is not directly suited for nominal or categorical predictors. To extend the IRA to categorical data, the method would need to be adapted to iterate across unique observed values of the focus predictor, rather than interpolated points. Second, the practical utility of IRA relies on the response variable having directly interpretable units of measurement. Although the method can technically be applied to bounded regression outputs, such as probabilities generated by a logistic regression model, the resulting IRA values would represent shifts in probability space, making direct physical interpretation less straightforward. Third, IRA values are intrinsically dataset-dependent and should not be compared across datasets with different predictor ranges, even if the underlying regression model is identical. Changing the predictor ranges will inherently alter the resulting IRA values. Fourth, IRA results may be affected by extreme values, especially for certain model types. IRA values should be interpreted with caution when predictors contain extreme values or when models are particularly sensitive to edge-case extrapolation. In such cases, restricting the analysis to a specific region of the range such as the $25^{th}$–$75^{th}$ inter-quartile range can mitigate outlier distortion and provide more reliable insights. The choice of range should therefore be guided by the study objectives and the region of practical or empirical relevance. Fifth, IRA does not explicitly model interaction effects as it evaluates predictors using a one-at-a-time perturbation strategy, holding background variables constant while varying the focus predictor. As a result, the framework does not explicitly capture or isolate multi-variable interaction terms. Therefore, IRA values should be interpreted strictly as measures of individual output sensitivity rather than indices of interaction strength. Finally, the optimal configuration of the framework's hyperparameters, such as the number of interpolated points ($M$) and the number of background observations ($K$), is not universal. These hyperparameters require further tuning to optimize IRA performance across different regression models and datasets.

## Conclusions

This study proposes the Impact Range Assessment (IRA), a model-agnostic sensitivity analysis method for regression models that quantifies a predictor's operational impact by evaluating the model's output shift across its observed range. Using synthetic linear and nonlinear datasets, IRA correctly separated relevant from irrelevant predictors across various hyperparameter settings. Furthermore, evaluations using the California Housing and Concrete Compressive Strength datasets across five machine learning models confirmed the effectiveness of the method at capturing architecture-specific sensitivity profiles.

Unlike traditional visual sensitivity tools such as PDP and ICE, the IRA method collapses complex multi-observation profiles into a single numeric value per predictor that can be ranked and compared directly without visual inspection. Additionally, while standard feature importance methods such as SHAP and permutation feature importance quantify a predictor's marginal contribution to predictive performance, IRA uniquely quantifies the potential scale of output variation over the predictor's full observed range.

Therefore, the IRA framework offers a valuable, computationally efficient, and complementary perspective to the traditional regression model interpretability toolkits that may be particularly useful for assessment of the practical sensitivity of the output to individual predictors. By translating abstract model sensitivities into tangible, unit-consistent measurements, the IRA provides actionable insights that are uniquely suited for process optimization, risk assessment, and decision-making across diverse scientific and industrial disciplines.

## Ethics statements

None.

## CRediT author statement

**Jihao You**: Conceptualization, Methodology, Writing - Original Draft, Writing - Review & Editing. **Dan Tulpan**: Supervision; Writing - Review & Editing. **Jiaojiao Diao**: Methodology, Writing - Review & Editing. **Jennifer L. Ellis**: Supervision; Funding acquisition; Project administration; Writing - Review & Editing.

## Acknowledgments

The authors would like to acknowledge the following funding agencies and institutions: Ontario Ministry of Agriculture, Food and Rural Affairs (OMAFRA) through the Ontario Agri-Food Innovation Alliance program in Canada, as well as the Natural Sciences and Engineering Research Council of Canada (NSERC Discovery program). The authors would like to further acknowledge Trouw Nutrition Canada for providing data for the real-world regression case study.

**Declaration of interests**

☒ The authors declare that they have no known competing financial interests or personal relationships that could have appeared to influence the work reported in this paper.

☐ The authors declare the following financial interests/personal relationships which may be considered as potential competing interests:

## Tables

Table 1. Single-execution Impact Range Assessment (IRA) for all predictors in the Linear Regression model on the synthetic linear dataset[1]

| Number of background samples[2] | Number of interpolated points[3] | $X_1$ | $X_2$* | $X_3$* | $X_4$* | $X_5$ | $X_6$* | $X_7$ | $X_8$* |
|---|---|---|---|---|---|---|---|---|---|
| 50 | 50 | 5.69 | 632.92 | 243.68 | 329.30 | 0.97 | 56.66 | 8.81 | 517.64 |
| | 100 | 5.69 | 632.92 | 243.68 | 329.30 | 0.97 | 56.66 | 8.81 | 517.64 |
| | 200 | 5.69 | 632.92 | 243.68 | 329.30 | 0.97 | 56.66 | 8.81 | 517.64 |
| | 500 | 5.69 | 632.92 | 243.68 | 329.30 | 0.97 | 56.66 | 8.81 | 517.64 |
| 100 | 50 | 5.69 | 632.92 | 243.68 | 329.30 | 0.97 | 56.66 | 8.81 | 517.64 |
| | 100 | 5.69 | 632.92 | 243.68 | 329.30 | 0.97 | 56.66 | 8.81 | 517.64 |
| | 200 | 5.69 | 632.92 | 243.68 | 329.30 | 0.97 | 56.66 | 8.81 | 517.64 |
| | 500 | 5.69 | 632.92 | 243.68 | 329.30 | 0.97 | 56.66 | 8.81 | 517.64 |
| 200 | 50 | 5.69 | 632.92 | 243.68 | 329.30 | 0.97 | 56.66 | 8.81 | 517.64 |
| | 100 | 5.69 | 632.92 | 243.68 | 329.30 | 0.97 | 56.66 | 8.81 | 517.64 |
| | 200 | 5.69 | 632.92 | 243.68 | 329.30 | 0.97 | 56.66 | 8.81 | 517.64 |
| | 500 | 5.69 | 632.92 | 243.68 | 329.30 | 0.97 | 56.66 | 8.81 | 517.64 |
| 500 | 50 | 5.69 | 632.92 | 243.68 | 329.30 | 0.97 | 56.66 | 8.81 | 517.64 |
| | 100 | 5.69 | 632.92 | 243.68 | 329.30 | 0.97 | 56.66 | 8.81 | 517.64 |
| | 200 | 5.69 | 632.92 | 243.68 | 329.30 | 0.97 | 56.66 | 8.81 | 517.64 |
| | 500 | 5.69 | 632.92 | 243.68 | 329.30 | 0.97 | 56.66 | 8.81 | 517.64 |

1. Predictors that contributed to the construction of the target variable Y are marked with an asterisk (*).
2. Number of background samples represents the number of background observations randomly drawn from the input dataset.
3. Number of interpolated points represents the number of values generated across the range of each focus predictor.

Table 2. Single-execution Impact Range Assessment (IRA) for all predictors in the Random Forest regressor on the synthetic nonlinear dataset[1]

| Number of background samples[2] | Number of interpolated points[3] | $X_1$* | $X_2$* | $X_3$* | $X_4$* | $X_5$* | $X_6$ | $X_7$ | $X_8$ |
|---|---|---|---|---|---|---|---|---|---|
| 50 | 50 | 7.76 | 8.88 | 5.33 | 12.05 | 6.66 | 2.50 | 2.33 | 2.43 |
| | 100 | 7.86 | 9.15 | 5.41 | 12.12 | 6.77 | 2.57 | 2.38 | 2.49 |
| | 200 | 7.96 | 9.24 | 5.44 | 12.17 | 6.79 | 2.59 | 2.43 | 2.50 |
| | 500 | 8.01 | 9.34 | 5.45 | 12.30 | 6.90 | 2.61 | 2.45 | 2.53 |
| 100 | 50 | 7.76 | 8.86 | 5.24 | 12.09 | 6.47 | 2.63 | 2.38 | 2.44 |
| | 100 | 7.84 | 9.14 | 5.32 | 12.20 | 6.57 | 2.69 | 2.43 | 2.50 |
| | 200 | 7.94 | 9.24 | 5.35 | 12.27 | 6.63 | 2.72 | 2.47 | 2.52 |
| | 500 | 8.00 | 9.34 | 5.37 | 12.36 | 6.72 | 2.74 | 2.49 | 2.54 |
| 200 | 50 | 8.05 | 8.75 | 5.34 | 11.97 | 6.58 | 2.62 | 2.42 | 2.51 |
| | 100 | 8.16 | 9.00 | 5.41 | 12.08 | 6.67 | 2.69 | 2.48 | 2.56 |
| | 200 | 8.26 | 9.10 | 5.45 | 12.15 | 6.75 | 2.71 | 2.51 | 2.58 |
| | 500 | 8.33 | 9.20 | 5.47 | 12.23 | 6.83 | 2.73 | 2.53 | 2.61 |
| 500 | 50 | 8.21 | 8.94 | 5.34 | 11.53 | 6.61 | 2.65 | 2.42 | 2.44 |
| | 100 | 8.31 | 9.18 | 5.41 | 11.63 | 6.71 | 2.71 | 2.48 | 2.49 |
| | 200 | 8.44 | 9.28 | 5.45 | 11.71 | 6.79 | 2.74 | 2.51 | 2.51 |
| | 500 | 8.50 | 9.38 | 5.48 | 11.78 | 6.87 | 2.75 | 2.53 | 2.53 |

1. Predictors that contributed to the construction of the target variable Y are marked with an asterisk (*).
2. Number of background samples represents the number of background observations randomly drawn from the input dataset.
3. Number of interpolated points represents the number of values generated across the range of each focus predictor.

Table 3. Repeated Impact Range Assessment (IRA) for all predictors in the Random Forest regressor on nonlinear data with different numbers of repeats[1]

| Number of repeats | $X_1$* | $X_2$* | $X_3$* | $X_4$* | $X_5$* | $X_6$ | $X_7$ | $X_8$ |
|---|---|---|---|---|---|---|---|---|
| 10 | 8.29 | 9.13 | 5.32 | 11.73 | 6.65 | 2.72 | 2.41 | 2.41 |
| CI (lower) | 7.94 | 8.58 | 5.14 | 11.49 | 6.28 | 2.55 | 2.30 | 2.28 |
| CI (upper) | 8.78 | 9.51 | 5.59 | 11.98 | 6.94 | 2.91 | 2.52 | 2.55 |
| 30 | 8.26 | 9.09 | 5.33 | 11.69 | 6.65 | 2.73 | 2.40 | 2.43 |
| CI (lower) | 7.90 | 8.54 | 5.04 | 11.30 | 6.24 | 2.58 | 2.30 | 2.30 |
| CI (upper) | 8.69 | 9.52 | 5.59 | 11.97 | 6.98 | 2.90 | 2.55 | 2.56 |
| 50 | 8.27 | 9.17 | 5.34 | 11.67 | 6.70 | 2.74 | 2.42 | 2.44 |
| CI (lower) | 7.83 | 8.61 | 5.05 | 11.28 | 6.28 | 2.60 | 2.28 | 2.31 |
| CI (upper) | 8.63 | 9.57 | 5.70 | 11.98 | 7.07 | 2.89 | 2.56 | 2.56 |
| 70 | 8.28 | 9.18 | 5.35 | 11.64 | 6.71 | 2.72 | 2.42 | 2.44 |
| CI (lower) | 7.85 | 8.62 | 5.06 | 11.23 | 6.33 | 2.55 | 2.28 | 2.30 |
| CI (upper) | 8.84 | 9.61 | 5.70 | 11.97 | 7.06 | 2.89 | 2.57 | 2.57 |
| 90 | 8.28 | 9.15 | 5.35 | 11.63 | 6.71 | 2.71 | 2.42 | 2.45 |
| CI (lower) | 7.84 | 8.52 | 5.05 | 11.25 | 6.36 | 2.53 | 2.28 | 2.29 |
| CI (upper) | 8.80 | 9.57 | 5.68 | 11.98 | 7.06 | 2.89 | 2.56 | 2.60 |

1. Predictors that contributed to the construction of the target variable Y are marked with an asterisk (*). 100 interpolated points for the focus predictors and 200 background observations are used within each repeat of the Impact Range Assessment (IRA) estimation process. The mean IRA values and 95% confidence intervals (CIs) are included in the table.

Table 4. Repeated Impact Range Assessment (IRA) on the California Housing dataset with five machine learning regressors[1]

| | *Median Income* | *House Age* | *Average Rooms* | *Average Bedrooms* | *Population* | *Average Occupancy* | *Latitude* | *Longitude* |
|---|---|---|---|---|---|---|---|---|
| GB | 3.08 | 0.26 | 1.31 | 0.32 | 0.12 | 1.20 | 2.14 | 2.13 |
| CI (lower) | 3.04 | 0.24 | 1.29 | 0.29 | 0.10 | 1.14 | 2.07 | 2.08 |
| CI (upper) | 3.11 | 0.28 | 1.34 | 0.35 | 0.14 | 1.26 | 2.22 | 2.18 |
| MLP | 2.71 | 0.58 | 34.48 | 5.51 | 1.70 | 2.72 | 3.58 | 5.61 |
| CI (lower) | 2.64 | 0.53 | 34.32 | 5.34 | 1.64 | 2.62 | 3.34 | 5.43 |
| CI (upper) | 2.78 | 0.64 | 34.63 | 5.64 | 1.77 | 2.84 | 3.85 | 5.82 |
| RF | 3.15 | 0.49 | 0.44 | 0.32 | 0.43 | 1.06 | 1.46 | 1.39 |
| CI (lower) | 3.10 | 0.43 | 0.41 | 0.30 | 0.39 | 0.94 | 1.37 | 1.32 |
| CI (upper) | 3.20 | 0.57 | 0.47 | 0.35 | 0.47 | 1.19 | 1.56 | 1.47 |
| RR | 6.33 | 0.48 | 15.13 | 21.75 | 0.14 | 4.71 | 3.96 | 4.36 |
| CI (lower) | 6.33 | 0.48 | 15.13 | 21.75 | 0.14 | 4.71 | 3.96 | 4.36 |
| CI (upper) | 6.33 | 0.48 | 15.13 | 21.75 | 0.14 | 4.71 | 3.96 | 4.36 |
| SVR | 3.77 | 0.69 | 1.63 | 1.64 | 1.24 | 3.10 | 2.87 | 4.06 |
| CI (lower) | 3.72 | 0.61 | 1.58 | 1.58 | 1.19 | 2.95 | 2.73 | 3.82 |
| CI (upper) | 3.82 | 0.76 | 1.69 | 1.72 | 1.30 | 3.23 | 3.00 | 4.25 |

GB: Gradient Boosting; MLP: Multilayer Perceptron; RF: Random Forest; RR: Ridge Regression; SVR: Support Vector Regression.

1. The unit of IRA values in this table is 100,000 USD. 100 interpolated points for the focus predictors and 200 background observations were used within each repeat of the 50-repeat Impact Range Assessment (IRA).

Table 5. Repeated Impact Range Assessment (IRA) on the Concrete Compressive Strength dataset with five machine learning regressors[1]

| | *Cement* | *Blast Furnace Slag* | *Fly Ash* | *Water* | *Superplasticizer* | *Coarse Aggregate* | *Fine Aggregate* | *Age* |
|---|---|---|---|---|---|---|---|---|
| GB | 37.99 | 15.28 | 2.34 | 16.28 | 5.78 | 9.05 | 6.64 | 36.47 |
| CI (lower) | 37.23 | 14.55 | 2.24 | 15.63 | 5.43 | 8.81 | 6.28 | 35.77 |
| CI (upper) | 38.76 | 15.94 | 2.45 | 16.83 | 6.20 | 9.36 | 7.08 | 37.15 |
| MLP | 53.61 | 39.38 | 17.63 | 15.66 | 13.05 | 11.88 | 11.86 | 36.45 |
| CI (lower) | 52.44 | 38.00 | 17.17 | 14.71 | 12.56 | 11.18 | 10.93 | 35.03 |
| CI (upper) | 54.96 | 40.64 | 18.30 | 16.56 | 13.62 | 12.61 | 12.91 | 37.72 |
| RF | 32.31 | 10.39 | 2.18 | 10.86 | 7.06 | 4.18 | 5.26 | 30.79 |
| CI (lower) | 31.10 | 9.61 | 1.97 | 9.92 | 6.65 | 3.86 | 4.82 | 29.91 |
| CI (upper) | 33.12 | 11.18 | 2.43 | 11.65 | 7.57 | 4.62 | 5.67 | 31.81 |
| RR | 51.72 | 36.60 | 17.12 | 19.52 | 9.37 | 5.67 | 7.26 | 41.49 |
| CI (lower) | 51.72 | 36.60 | 17.12 | 19.52 | 9.37 | 5.67 | 7.26 | 41.49 |
| CI (upper) | 51.72 | 36.60 | 17.12 | 19.52 | 9.37 | 5.67 | 7.26 | 41.49 |
| SVR | 18.48 | 9.27 | 5.90 | 13.17 | 10.46 | 6.59 | 8.56 | 21.27 |
| CI (lower) | 17.73 | 8.89 | 5.41 | 12.58 | 9.89 | 6.04 | 7.99 | 20.48 |
| CI (upper) | 19.23 | 9.67 | 6.36 | 13.64 | 11.18 | 7.16 | 9.15 | 22.02 |

GB: Gradient Boosting; MLP: Multilayer Perceptron; RF: Random Forest; RR: Ridge Regression; SVR: Support Vector Regression.

1. The unit of IRA values in this table is MPa. 100 interpolated points for the focus predictors and 200 background observations were used within each repeat of the 50-repeat Impact Range Assessment (IRA).

Table 6. Repeated Impact Range Assessment (IRA) analysis for Pellet Durability Index (PDI) prediction using multiple linear regression model on the feed mill data

| Predictor | Mean | CI (lower) | CI (upper) |
|---|---|---|---|
| *Fat Content* | 5.51 | 5.44 | 5.57 |
| *ADF Content* | 4.65 | 4.61 | 4.68 |
| *Indoor Humidity (Pelletizer)* | 3.19 | 3.16 | 3.23 |
| *Dehydrated Bakery Meal* | 3.07 | 3.04 | 3.09 |
| *Amino Acids* | 2.80 | 2.77 | 2.83 |
| *Expanding Temperature* | 2.32 | 2.30 | 2.34 |
| *Processing Aid Water* | 2.00 | 1.98 | 2.02 |
| *Ambient Humidity* | 1.66 | 1.64 | 1.67 |
| *Cumulative Production* | 1.58 | 1.56 | 1.60 |

1. 100 interpolated points for the focus predictors and 200 background observations were used within each repeat of the 50-repeat Impact Range Assessment (IRA).

Figure 1. Pseudocode for the Impact Range Assessment (IRA) value estimation.

```
Require: Input dataset S (of size n observations × p continuous numeric predictors), regression model model, number of interpolated points M, number of random samples K
Ensure: Estimated Impact Range Assessment (IRA) value for each predictor
for each predictor i in S do
    Generate M values for predictor i from its minimum to maximum (e.g., via linear interpolation)
    Sample K full observations from S as background samples
    Duplicate each of the K background samples M times to obtain K × M total modified observations
    for each background sample do
        Replace predictor i with the M generated values (keep other predictors fixed)
        Use model to predict outputs for all modified observations
        Calculate range as max(ŷ) − min(ŷ) where ŷ is the model output
    end for
    Compute the Impact Range Assessment (IRA) value by averaging the range across all background samples
    Store (predictor i, Impact Range Assessment (IRA) value))
end for
return Impact Range Assessment (IRA) values for all predictors
```

**Supplementary Materials**

Table S1 Statistics description of the synthetic linear dataset.

| Variable[1] | Mean | SD | Minimum | Median | Maximum |
|---|---|---|---|---|---|
| $X_1$ | -0.002 | 1.003 | -3.655 | -0.016 | 3.603 |
| $X_2$ | -0.015 | 1.005 | -3.837 | 0.007 | 3.728 |
| $X_3$ | -0.010 | 0.989 | -3.368 | -0.020 | 3.853 |
| $X_4$ | 0.027 | 1.008 | -3.376 | 0.035 | 3.611 |
| $X_5$ | -0.008 | 0.993 | -3.233 | -0.009 | 3.926 |
| $X_6$ | 0.013 | 1.003 | -3.688 | 0.018 | 4.479 |
| $X_7$ | -0.009 | 0.985 | -4.466 | -0.005 | 3.335 |
| $X_8$ | -0.013 | 1.014 | -3.922 | -0.012 | 3.378 |
| $Y$ | 99.427 | 134.182 | -362.102 | 99.062 | 546.922 |

1. The synthetic linear dataset was generated using the *make_regression()* function from the *scikit-learn* library. $X_2$, $X_3$, $X_4$, $X_6$ and $X_8$ are relevant predictors, while $X_1$, $X_5$ and $X_7$ are irrelevant predictors. $Y$ is the response variable. The true coefficients generated by the *make_regression()* function for $X_2$, $X_3$, $X_4$, $X_6$ and $X_8$ were 83.17, 34.54, 46.40, 5.06, and 70.54, respectively.

Table S2 Statistics description of the synthetic nonlinear dataset.

| Variable[1] | Mean | SD | Minimum | Median | Maximum |
|---|---|---|---|---|---|
| $X_1$ | 0.502 | 0.284 | 0.000 | 0.508 | 1.000 |
| $X_2$ | 0.493 | 0.290 | 0.000 | 0.490 | 1.000 |
| $X_3$ | 0.497 | 0.287 | 0.000 | 0.496 | 1.000 |
| $X_4$ | 0.503 | 0.288 | 0.000 | 0.501 | 1.000 |
| $X_5$ | 0.498 | 0.288 | 0.000 | 0.496 | 1.000 |
| $X_6$ | 0.500 | 0.289 | 0.000 | 0.502 | 1.000 |
| $X_7$ | 0.503 | 0.288 | 0.000 | 0.504 | 0.999 |
| $X_8$ | 0.500 | 0.291 | 0.000 | 0.504 | 1.000 |
| $Y$ | 14.264 | 7.834 | -12.124 | 14.278 | 44.209 |

1. The synthetic nonlinear dataset was generated using the *make_friedman1()* function from the *scikit-learn* library. $X_1$, $X_2$, $X_3$, $X_4$ and $X_5$ are relevant variables that were used to compute the response variable $Y$, whereas $X_6$, $X_7$ and $X_8$ are irrelevant variables.

Table S3 Statistics description of the California Housing dataset.

| Variable | Mean | SD | Minimum | Median | Maximum |
|---|---|---|---|---|---|
| *Median Income* | 3.871 | 1.900 | 0.500 | 3.535 | 15.000 |
| *House Age* | 28.639 | 12.586 | 1.000 | 29.000 | 52.000 |
| *Average Rooms* | 5.429 | 2.474 | 0.846 | 5.229 | 141.909 |
| *Average Bedrooms* | 1.097 | 0.474 | 0.333 | 1.049 | 34.067 |
| *Population* | 1425.477 | 1132.462 | 3.000 | 1166.000 | 35682.000 |
| *Average Occupancy* | 3.071 | 10.386 | 0.692 | 2.818 | 1243.333 |
| *Latitude* | 35.632 | 2.136 | 32.540 | 34.260 | 41.950 |
| *Longitude* | -119.570 | 2.004 | -124.350 | -118.490 | -114.310 |
| *Median House Value* | 2.069 | 1.154 | 0.150 | 1.797 | 5.000 |

Table S4 Statistics description of the Concrete Compressive Strength dataset.

| Variable | Mean | SD | Minimum | Median | Maximum |
|---|---|---|---|---|---|
| *Cement* | 281.166 | 104.507 | 102.000 | 272.900 | 540.000 |
| *Blast Furnace Slag* | 73.895 | 86.279 | 0.000 | 22.000 | 359.400 |
| *Fly Ash* | 54.187 | 63.996 | 0.000 | 0.000 | 200.100 |
| Water | 181.566 | 21.356 | 121.750 | 185.000 | 247.000 |
| *Superplasticizer* | 6.203 | 5.973 | 0.000 | 6.350 | 32.200 |
| *Coarse Aggregate* | 972.919 | 77.754 | 801.000 | 968.000 | 1145.000 |
| *Fine Aggregate* | 773.579 | 80.175 | 594.000 | 779.510 | 992.600 |
| *Age* | 45.662 | 63.170 | 1.000 | 28.000 | 365.000 |
| *Compressive Strength* | 35.818 | 16.706 | 2.332 | 34.443 | 82.599 |

Table S5 Statistics description of the feed mill data and the estimated coefficients of the multiple linear regression model in the Pellet Durability Index (PDI) prediction study.

| Variable | Mean | SD | Minimum | Median | Maximum | Coefficient[2] |
|---|---|---|---|---|---|---|
| *Amino Acids* | 0.4 | 0.2 | 0.0 | 0.4 | 1.0 | $4.42 \times 10^8$ |
| *ADF Content* | 3.4 | 1.2 | 1.7 | 2.9 | 7.2 | $1.41 \times 10^8$ |
| *Dehydrated Bakery Meal* | 6.1 | 5.5 | 0.0 | 5.9 | 16.5 | $3.06 \times 10^7$ |
| *Indoor Humidity (Pelletizer)* | 28.1 | 9.8 | 10.5 | 27.3 | 53.8 | $1.20\times 10^7$ |
| *Expanding Temperature* | 92.1 | 6.9 | 62.5 | 91.8 | 111.7 | $7.50 \times 10^6$ |
| *Cumulative Production* | 20687.7 | 13415.3 | 55.4 | 18602.4 | 47953.9 | $-5.29 \times 10^3$ |
| *Ambient Humidity* | 65.3 | 13.1 | 22.5 | 66.5 | 91.8 | $-3.88 \times 10^6$ |
| *Fat Content* | 3.7 | 0.8 | 2.1 | 3.7 | 7.5 | $-1.55 \times 10^8$ |
| *Processing Aid Water* | 0.9 | 0.2 | 0.0 | 0.8 | 1.5 | $-2.17 \times 10^8$ |

1. The statistics description of the data refers to the training set used in the study.

2. Coefficient refers to the parameters of the multiple linear regression model for calculating the transformed Pellet Durability Index (PDI). The intercept of the multiple linear regression model is $2.14 \times 10^9$. The large coefficient and intercept values result from applying Box-Cox transformation to PDI (response variable). The Box-Cox parameter ($\lambda$) estimated in the study was 5.18.

Table S6. Runtime of the single-execution Impact Range Assessment (IRA) across synthetic and benchmark dataset examples[1].

| Dataset | Model | Runtime (s) |
|---|---|---|
| Synthetic linear dataset[2] | LR | 0.065 ± 0.011 |
| Synthetic nonlinear dataset[3] | RF | 0.751 ± 0.026 |
| | GB | 0.183 ± 0.009 |
| | MLP | 0.156 ± 0.015 |
| California Housing dataset[4] | RF | 0.849 ± 0.016 |
| | RR | 0.073 ± 0.009 |
| | SVR | 167.182 ± 3.559 |
| | GB | 0.182 ± 0.008 |
| | MLP | 0.151 ± 0.009 |
| Concrete Compressive Strength dataset[5] | RF | 0.598 ± 0.085 |
| | RR | 0.096 ± 0.023 |
| | SVR | 10.612 ± 0.117 |

GB: Gradient Boosting; LR: Linear Regression; MLP: Multilayer Perceptron; RF: Random Forest; RR: Ridge Regression; SVR: Support Vector Regression.

1. All runtimes were measured on a machine with an AMD Ryzen 7 4800H processor (2.90 GHz) and 16.0 GB of installed RAM, using 100 interpolated points and 200 background observations. Each model–dataset combination was evaluated in 30 consecutive single-execution IRA runs. Reported values are mean ± standard deviation across all runs. Repeating the 30-run benchmark may yield slightly different estimates because of normal variation in system load and execution conditions.

2. The synthetic linear dataset was generated using the *make_regression()* function from the *scikit-learn* library, containing 5000 observations and 8 predictors.

3. The synthetic nonlinear dataset was generated using the *make_friedman1()* function from the *scikit-learn* library, containing 5000 observations and 8 predictors.

4. The California Housing dataset contained 20640 observations and 8 predictors.

5. The Concrete Compressive Strength dataset contained 1030 observations and 8 predictors.

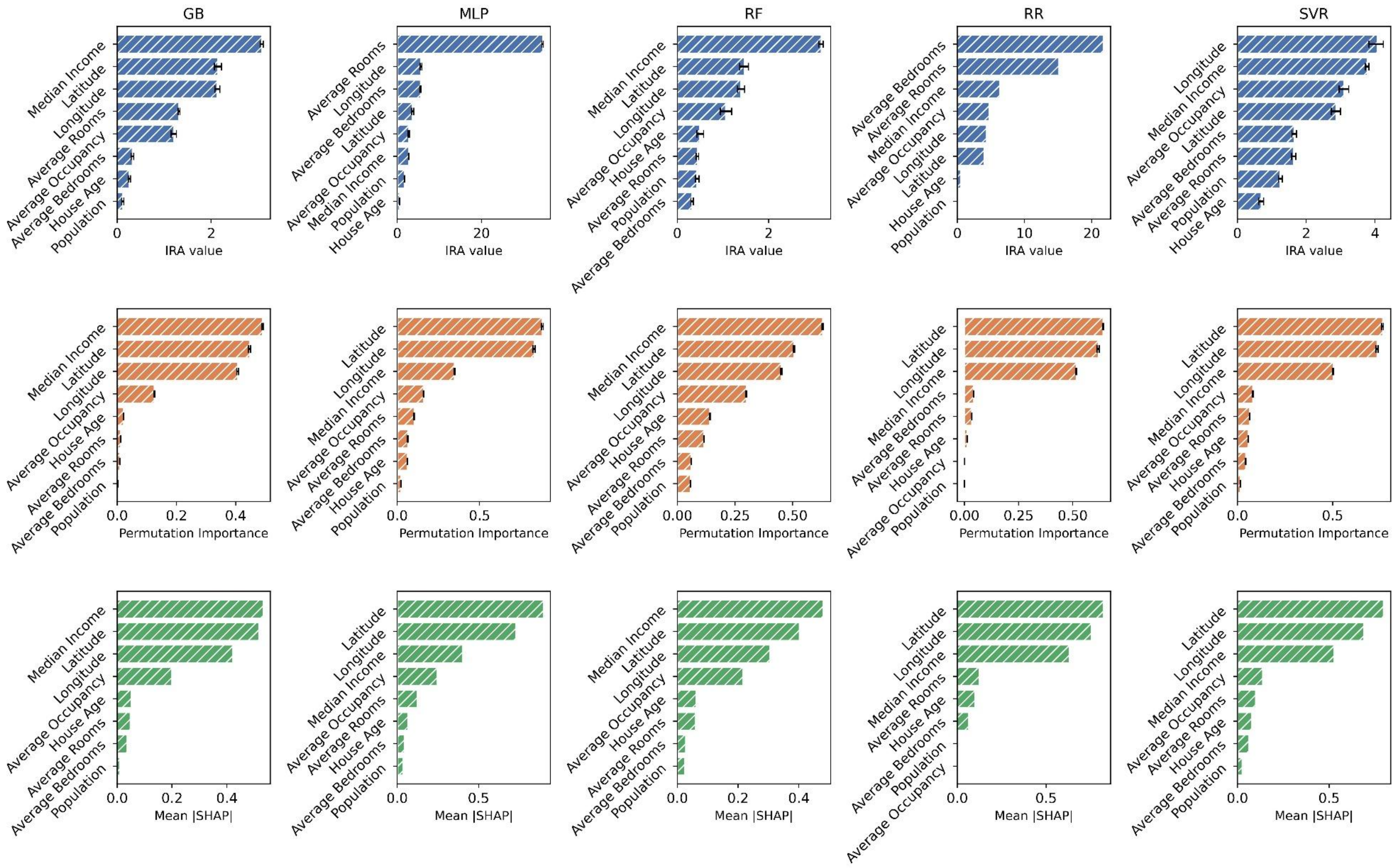

Figure S1. Comparison of predictor ranking based on repeated Impact Range Assessment (IRA), permutation importance, and SHAP value for five machine learning regressors on the California Housing dataset.

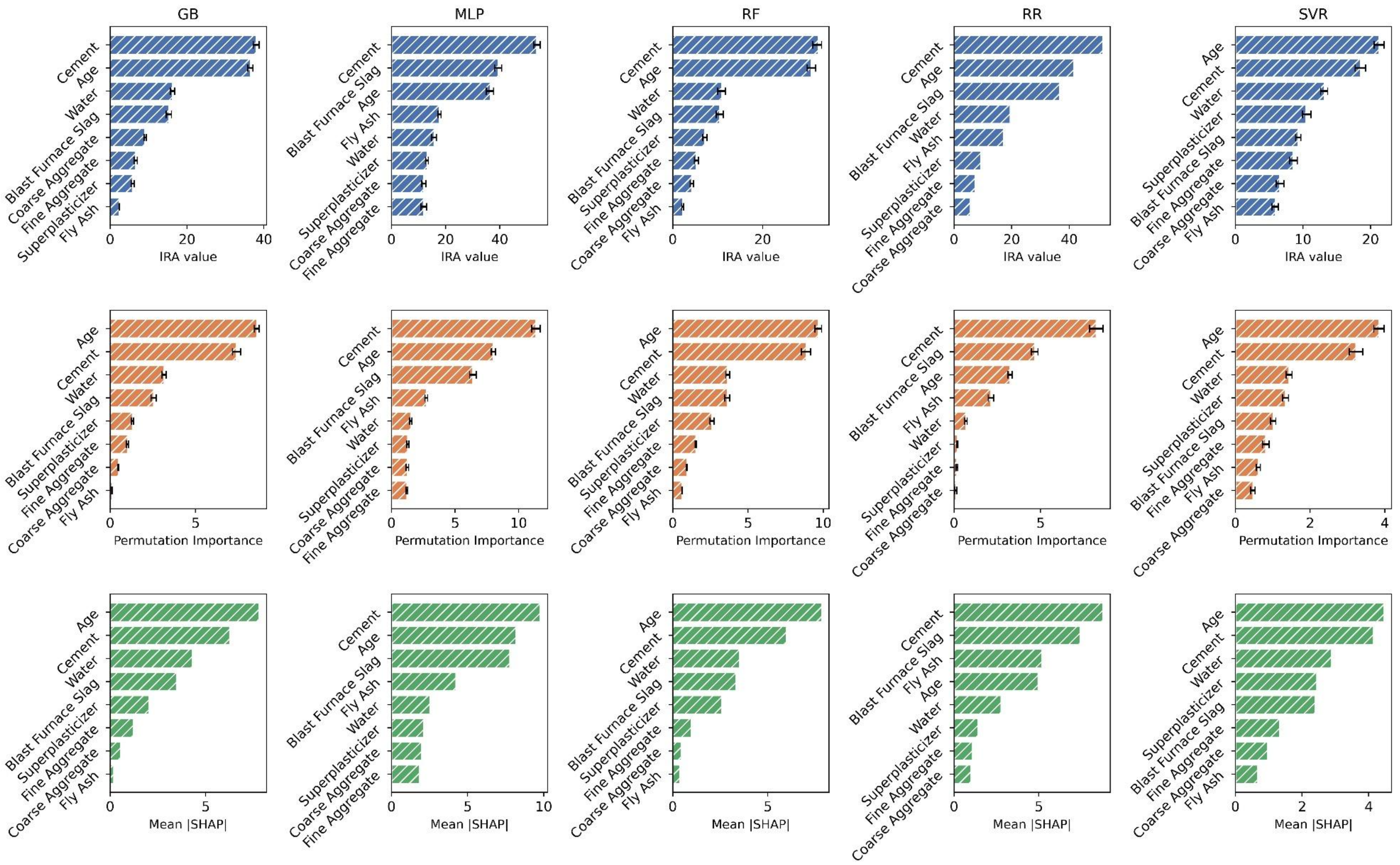

Figure S2. Comparison of predictor ranking based on repeated Impact Range Assessment (IRA), permutation importance, and SHAP value for five machine learning regressors on the Concrete Compressive Strength dataset.

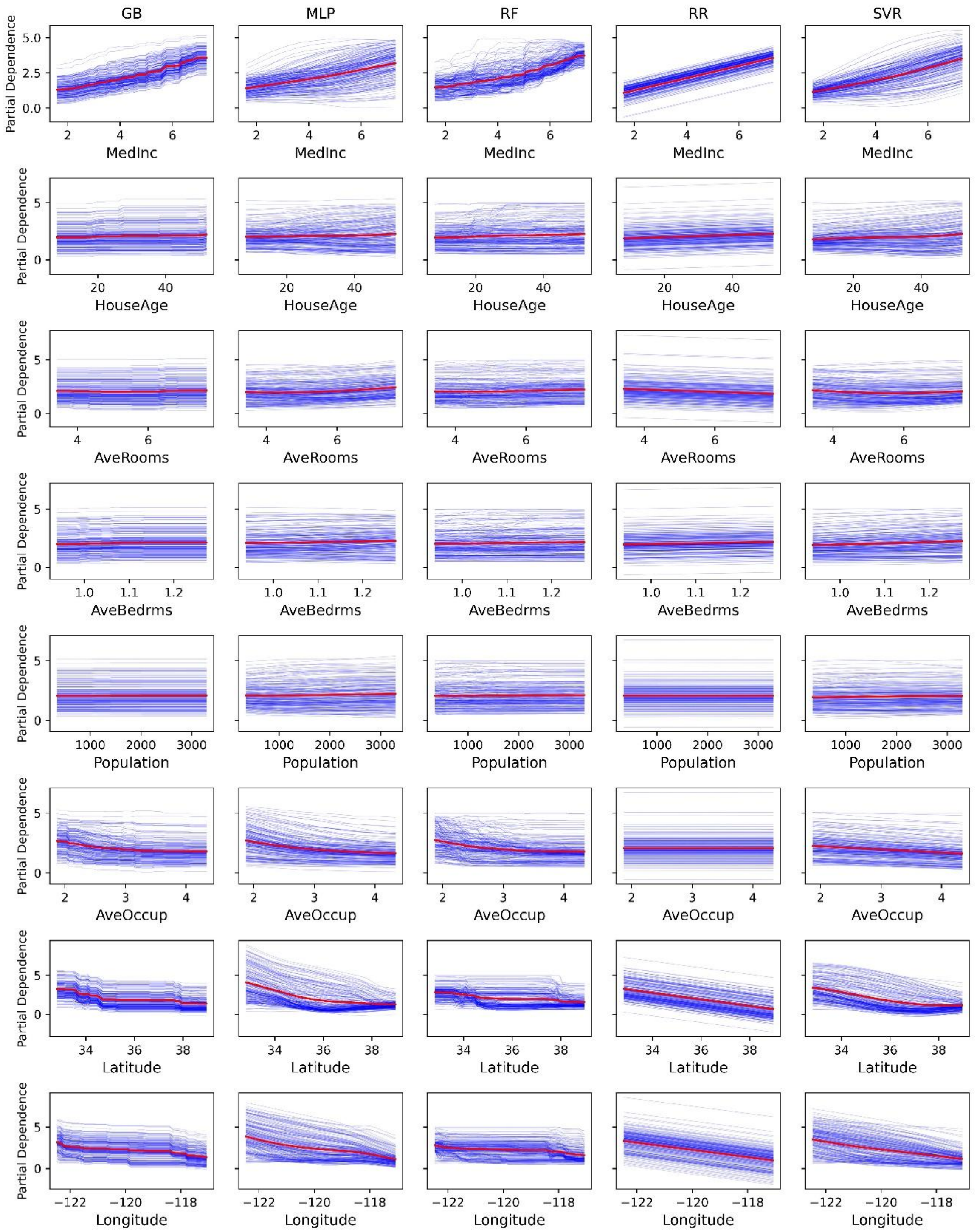

Figure S3. Partial Dependence Plots (PDP) and Individual Conditional Expectation (ICE) plots for five machine learning regressors trained on the California Housing dataset. Each row represents a predictor and each column a model; red lines show the partial dependence computed at 50 grid points, and blue lines show the individual conditional expectations for 200 randomly sampled observations.

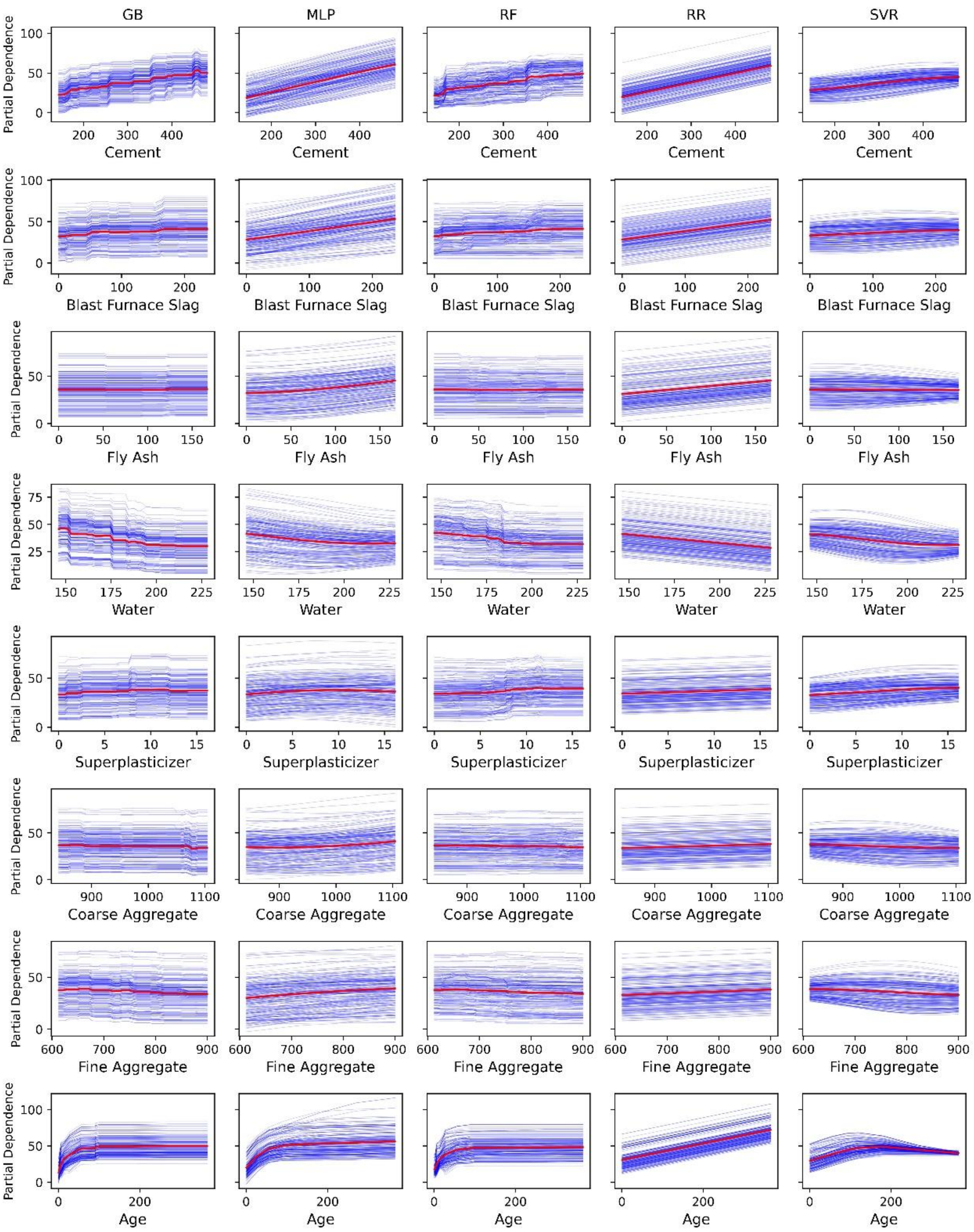

Figure S4. Partial Dependence Plots (PDP) and Individual Conditional Expectation (ICE) plots for five machine learning regressors trained on the Concrete Compressive Strength dataset.

Each row represents a predictor and each column a model; red lines show the partial dependence computed at 50 grid points, and blue lines show the individual conditional expectations for 200 randomly sampled observations.